\documentclass[twocolumn,showpacs,preprintnumbers,amsmath,amssymb]{revtex4}

\usepackage{graphicx}
\usepackage{dcolumn}
\usepackage{bm}

\begin{document}
\title{Diffusion-limited-aggregation on a directed small world network}
\author{Jie Ren$^{1}$}
\author{Wen-Xu Wang$^{2}$}
\author{Gang Yan$^{3}$}
\author{Bing-Hong Wang$^{2}$}
\email{bhwang@ustc.edu.cn}
\affiliation{$^{1}$Department of Physics,
\\$^{2}$Department of Modern Physics,
\\$^{3}$Department of Electronic Science and Technology, \\University of
Science and Technology of China, Hefei, 230026, PR China }

\date{\today}

\begin{abstract}
For real world systems, nonuniform medium is ubiquitous. Therefore,
we investigate the diffusion-limited-aggregation process on a two
dimensional directed small-world network instead of regular lattice.
The network structure is established by rewiring connections on the
two dimensional directed lattice. Those rewired edges are controlled
by two parameters $\theta$ and $m$, which characterize the spatial
length and the density of the long-range connections, respectively.
Simulations show that there exists a maximum value of the fractal
dimension when $\theta$ equals zero. Interestingly, we find that the
symmetry of the aggregation pattern is broken when rewired
connections are long enough, which may be an explanation for the
formation of asymmetrical fractal in nature. Then, we perform
multifractal analysis on the patterns further.
\end{abstract}

\maketitle
\section{INTRODUCTION}
Nonequilibrium growth models leading naturally to self-organized
fractal patterns, the structure of which strongly depends on the
dynamics of the growth process, are of continuing interest due to
their relevance for many important fields\cite{Meakin}.
Diffusion-limited -aggregation (DLA)\cite{Witten} is probably the
most remarkable growth model for pattern formation. This model
generates complex and mysterious fractal
structures\cite{Vicsek,Stanley}, which seem to be generated as well
in natural systems in which growth is controlled by diffusive,
including dielectric breakdown\cite{dielectric}, electrochemical
deposition\cite{electrochemical}, colloid aggregation\cite{Kolb},
film growth\cite{Elam}, viscous fingering\cite{viscous}, Laplacian
flow\cite{Laplacian} etc.

In the original DLA model\cite{Witten}, particles released at a
point distant from the cluster execute random walks until they find
a nearest neighbor site of the cluster and irreversibly stick at
this site. However, for many physical processes, the medium is
nonuniform so that the probability of jumping from site $i$ to site
$j$ is usually not equal to that from $j$ to $i$. Moreover, except
nearestneighboring jumps, there also exist some nonlocal jumps
through which the particle can move to a distant site at a step. One
example is the diffusion of adatoms on metal surfaces, in which the
long jumps play a dominating role\cite{Long1,Long2}. A bunch of
defects or impurities in the substrate may also play the part of the
long-jump path in the case of weak absorbate-substrate
interaction\cite{Long3}, which is important for the thin-film
growth, heterogeneous catalysis, and oxidation. Hence, the
traditional undirected regular lattice will miss important
information of medium and it is unconformity to characterize the
actual DLA process. This thus calls for the use of network structure
with directed long-range connections.

On the other hand, complex networks have recently attracted an
increasing interest among physicists\cite{Review1,Review2,Review3}.
In particular, small-world (SW) networks, introduced by Watts and
Strogatz\cite{WS}, have been extensively studied because they
constitute an interesting attempt to translate the complex topology
of social, biological, and physical networks into a simple model.
Two dimensional (2D) SW networks result from randomly rewiring a
fraction of links of a 2D regular lattice. Several dynamical models
have recently been studied in order to understand the effect of SW
topology on classical systems such as the Ising model\cite{Ising},
the spread of epidemics\cite{Epidemic}, aggregation\cite{Huang},
random walks\cite{RW}, etc. Such models are expected to capture the
essential features of the complicated processes taking place on real
networks.

In this paper, we investigate the DLA process on a 2D directed SW
network, in which the directed links correspond to the directed
irreversibly jumps and the node is regarded as the lattice point of
the real space, respectively.

\section{THE MODEL}
In order to construct the directed SW network, we start from a 2D
square lattice of size $L\times L$ consisting of sites linked to
their four nearest neighbors by both outgoing and incoming links, as
shown in Fig. 1. Then, we reconnect nearest-neighbor outgoing links
to a different site chosen at random with the probability
\begin{equation}
p(r)\sim e^{-\theta r},
\end{equation}
where $r$ is the lattice distance between the two selected sites and
$\theta(\theta\in[-1,1])$ is the clustering exponent which
characterizes the localized degree of the long-range links in the
lattice. The formula corresponds to the diffusivity represented by
Arrhenius relation as usual in surface
science\cite{Long1,Long2,Long3}. The reconstructing process is
repeated until $m$, the number of long-range rewiring connections,
reaches a desired value. Note that by performing this procedure
every site will have exactly four fixed outgoing links and a random
number of incoming links. When the clustering exponent $\theta$=0,
we have a uniform distribution over long-range connections, and the
present model reduces to the basic 2D directed SW
network\cite{directed}. As $\theta\rightarrow1$
($\theta\rightarrow1$ denotes $\theta$ tends to 1), the long-range
links of a site become more and more local in its vicinity. In
reverse, as $\theta\rightarrow-1$, the long-range rewiring outgoing
links are in favor of the farther sites. Thus, the clustering
exponent $\theta$ serves as a structural parameter controlling the
spatial length of the long-range connections.

\begin{figure}
\scalebox{0.4}[0.4]{\includegraphics{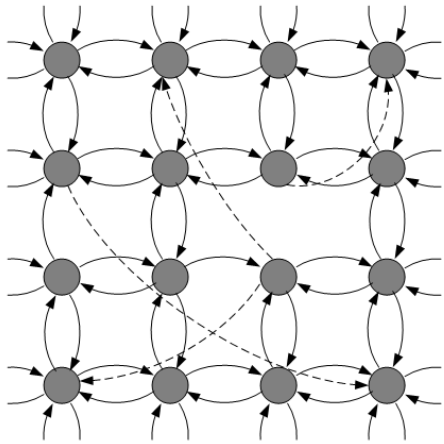}}
\caption{\label{fig:epsart} Sketch map of a directed 2D SW lattice.
Dotted lines represent rewired connections. Arrows indicate the
direction of the corresponding connection.}
\end{figure}

Based on the directed 2D SW network as constructed above, we have
performed extensive numerical simulations for the DLA with size of
the reconstructed lattice $L$=1000 with number of particles
$N$=10000. Staring from an immobile seed at the center of the
lattice, a particle is released at a random position where is depart
from the outer radius of the growing pattern. Then the particle
jumps along the direction from the current site to one of its linked
sites which are not occupied by the growing pattern, with equal
probability step by step. At last, the particle irreversibly sticks
at the nearest neighbor site of the growing pattern in terms of the
physical distance and the pattern will grow gradually. To reduce the
effect of fluctuation, the calculated result is taken average over
10 different network realizations and 10 independent runs for each
network configuration for each set of parameters $(\theta, m)$.

\section{SIMULATION RESULT AND DISCUSSION}
\begin{figure}
\scalebox{0.5}[0.5]{\includegraphics{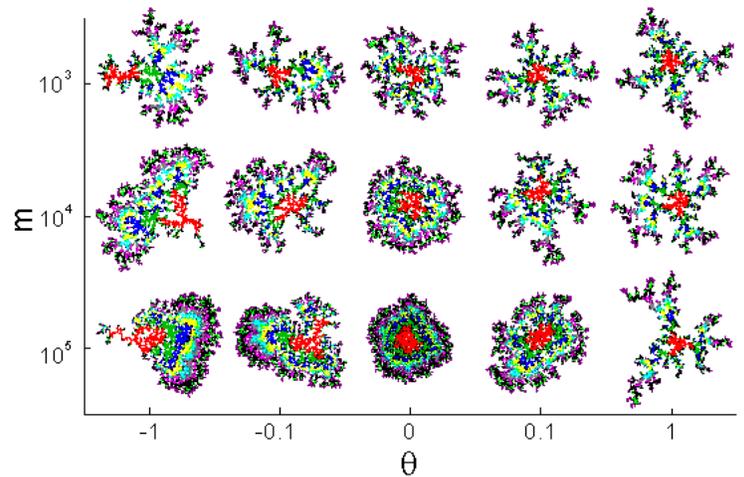}} \caption{The
simulation patterns as a function of the clustering exponent for the
number of long-range connections $m$=1000, 10000 and 100000
respectively. Each color represents 1000 particles in order.}
\end{figure}

Fig 2 illustrates the patterns of DLA which exhibit rich behaviors
for different parameters $\theta$ and $m$. For each $\theta$, it can
be seen that with the increase of $m$ the patterns of DLA become
thicker and denser, however, which is not obvious for large
$\theta$, approximately 1. For each $m$, the pattern is nearly the
most dense when $\theta$=0. While, it gets thin and sparse when
$\theta$ departs from 0 tending to 1 or $-1$. However, it is
astonishing that the symmetry of the aggregation pattern is markedly
broken while $\theta<0$, which is more obvious as $\theta$ tends to
$-1$. To quantify the patterns of DLA, we calculate the fractal
dimensions $D_0$ of the DLA by box-counting
method\cite{Vicsek,Stanley}, which are shown in Fig. 3. It is clear
that there exists a maximum value of $D_0$ when $\theta$ equals 0
for each $m$. It can be seen that the more $\theta$ departs from 0,
the more $D_0$ decreases. Moreover, it is found that $D_0$ decreases
more fast when $\theta\rightarrow1$ than $\theta\rightarrow-1$.

\begin{figure}
\scalebox{0.85}[0.85]{\includegraphics{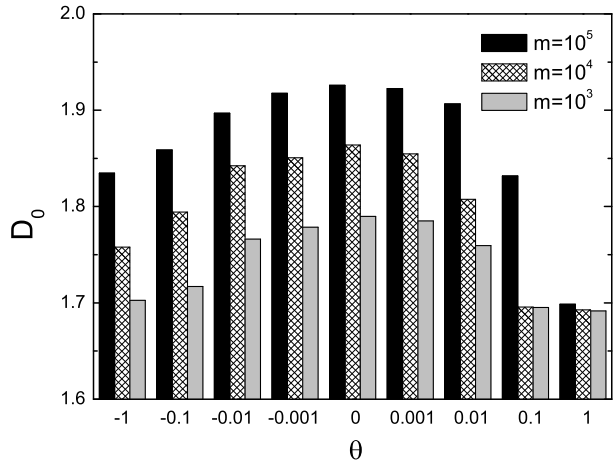}}
\caption{\label{fig:epsart} The fractal dimension $D_0$ of patterns
as a function of the clustering exponent $\theta$ for several $m$.}
\end{figure}

It is well-known that the special randomly branching, open structure
of a DLA pattern results from the effects of screening, which is
manifested through the fact that the tips of most advanced branches
capture the incoming diffusing particles most effectively. In the
present work, due to the long-range connections, particles can jump
directly to distant sites, including the traditionally completely
screened deep fjord. The nonlocal connections effectively weaken the
screening effect so that the pattern of aggregate becomes compact
and the fractal dimension $D_0$ increases with $m$ increasing. On
the other hand, the clustering exponent $\theta$, which restricts
the spatial length of the long-range connection, affects the
morphology of the aggregate and the fractal dimension $D_0$. As
$\theta\rightarrow1$, the long-range links are restricted more and
more local in its vicinity so that the capacity of weakening the
screening effect vanishes gradually. Finally, $D_0$ does not vary
and the morphology of the aggregate seems like the original DLA
pattern, as shown in Fig. 2 and Fig. 3. When $\theta$=0, the spatial
lengths of the long-range connections are entire random, neither too
distant nor too local and they have a uniform distribution. Then,
due to intensive weakening for the screening effect, the random
particle has the chance to appear on arbitrary sites on the
underlying network so that the pattern becomes thick and compact,
corresponding to increase of $D_0$. However, as
$\theta\rightarrow-1$, the long-range links tend to the sites as
distant as possible and the irreversible jumps along directed links
break the symmetry of dynamics. Thus, small fluctuations are
enhanced, and this instability together with the randomness inherent
in the model leads to a complex asymmetrical behavior.

However, the fractal dimension $D_0$ is a rough description because
the pattern becomes asymmetric while $\theta\rightarrow-1$. So, we
have performed the multifractal analyse\cite{multi1,multi2} here to
see more details. It should be noted that our measurements concern
the pattern itself other than its harmonic measure\cite{multi}.

Further characterization of the DLA patterns can be achieved by
determining the generalized fractal dimensions $D_q$. Cover the
pattern with a grid of square boxes of size $\varepsilon$ and define
$P_i(\varepsilon)$ to be the relative portion of the pattern in cell
$i$, and define $N(\varepsilon)$ to be the total number of boxes of
size $\varepsilon$ needed to cover the whole pattern. The relative
portion $P_i(\varepsilon)$ can be described as multifractal as:
\begin{equation}
P_i(\varepsilon)\sim\varepsilon^{\alpha},
\end{equation}
\begin{equation}
N_\alpha(\varepsilon)\sim \varepsilon^{-f(\alpha)},
\end{equation}
where $\alpha$ is the singularity, $N_\alpha(\varepsilon)$ the
number of small squares of relative size $\varepsilon$ with the same
singularity, and $f(\alpha)$ is the fractal dimension.

\begin{figure}
\scalebox{0.85}[0.85]{\includegraphics{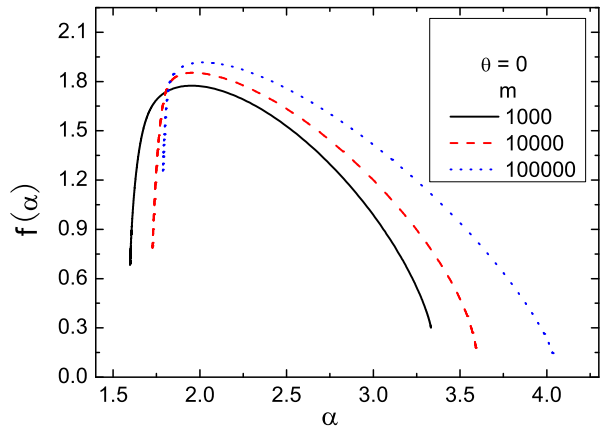}}
\caption{\label{fig:epsart} The multifractal spectra $f(\alpha)$ of
the distribution of the patterns, with various number of long-range
connections, $m$=1000, 10000, 100000, respectively.}
\end{figure}

To describe quantitatively the subtle geometrical feature of the
pattern, the fractal dimension $f(\alpha)$ can be obtain from the
partition function $\chi_q$:
\begin{equation}
\chi_q=\sum_{i}^{N(\varepsilon)}P_i^q(\varepsilon),
\end{equation}
and its power law of $\varepsilon$,
\begin{equation}
\tau_q=\lim_{\varepsilon\rightarrow0}\frac{\ln\chi_q}{\ln\varepsilon},
\end{equation}
where $q$ is the moment order and $\tau_q$ the index of the pow law.
The generalized fractal dimension is defined as:
\begin{equation}
D_q=\frac{\tau_q}{q-1},
\end{equation}
Then, $(\alpha, f(\alpha))$ can be obtained from $(q, D_q)$ by
performing the Legendre transformation:
\begin{equation}
\alpha=\frac{d}{dq}[(q-1)D_q],
\end{equation}
\begin{equation}
f(\alpha)=\alpha q-(q-1)D_q,
\end{equation}
In our calculation, the moment order $q$ is taken for -30 to 30.

\begin{figure}
\scalebox{0.85}[0.85]{\includegraphics{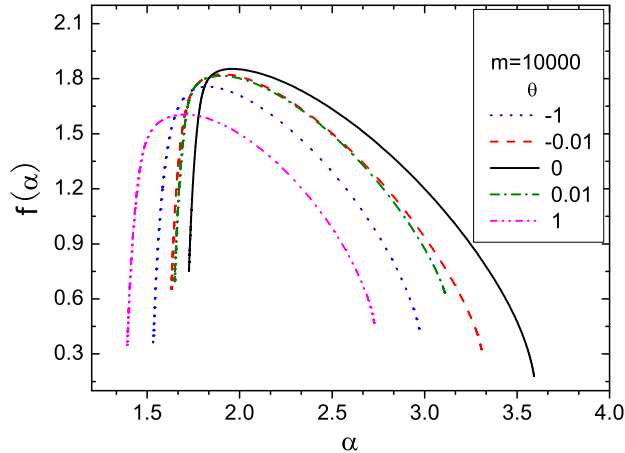}}
\caption{\label{fig:epsart} The multifractal spectra with several
value of the cluster exponent $\theta$, -1, -0.01, 0, 0.01, 1, for
$m$=10000.}
\end{figure}

We have calculated the multifractal spectra $f(\alpha)$ of the
distribution of the patterns, with various number of long-range
connections, $m$=1000, 10000, 100000, respectively, for a original
directed 2D SW lattice, $\theta$=0. Figure 4 shows the result. It
can be seen that the curve becomes higher and the range of
singularity $\alpha$ becomes wider with increasing the number of the
long-range connections. In Table I, the multifractal parameters of
the distribution are listed. The multifractal spectrum can be used
to provide more information about the subtle geometrical difference,
because of the $\alpha_{max}$ and $\alpha_{min}$ connecting with the
smallest probability and the largest probability of the spatial
distribution [show by Eq .(2)]. The result (Table I) show that
$\alpha_{max}$ and $\alpha_{min}$ both increase with increasing $m$,
while $\Delta\alpha$=$\alpha_{max}-\alpha_{min}$ also increases,
indicating that the pattern becomes less irregular, less nonuniform,
and more dense. Moreover, Fig. 5 illustrates that the multifractal
curves with several value of the cluster exponent $\theta$, -1,
-0.01, 0, 0.01, 1, for a directed 2D SW lattice, $m$=10000. More
quantitative details can be seen in Table II. It illustrates that
the range of $\alpha$ is the broadest and the curve is the maximal
when $\theta$=0, suggesting the pattern is the most dense, compact
and regular, which corresponds to Fig. 2 and Fig. 3 showed above.

\begin{center}
 {TABLE I: Some multifractal parameters of Figure 4.}
  \begin{tabular}{cccc}\hline\hline
   $m$   & 1000 & 10000 & 100000 \\ \hline
   $\alpha_{min}$   & 1.598 & 1.726 & 1.786 \\
   $\alpha_{max}$   & 3.332 & 3.591 & 4.034 \\
   $\Delta\alpha=\alpha_{max}-\alpha_{min}$   & 1.734 & 1.865 & 2.248 \\
   $f(\alpha_{min})$   & 0.684 & 0.752 & 1.251 \\
   $f(\alpha_{max})$   & 0.303 & 0.180 & 0.147 \\
   $\Delta f=f(\alpha_{min})-f(\alpha_{max})$   & 0.381 & 0.572 & 1.104 \\
   \hline\hline
 \end{tabular}
\end{center}

\begin{center}
 {TABLE II: Some multifractal parameters of Figure 5.}
  \begin{tabular}{cccccc}\hline\hline
   $\theta$   & -1 & -0.01 & 0 & 0.01 & 1 \\ \hline
   $\alpha_{min}$   & 1.532 & 1.630 & 1.726 & 1.648 & 1.392 \\
   $\alpha_{max}$   & 2.969 & 3.307 & 3.591 & 3.109 & 2.727 \\
   $\Delta\alpha=\alpha_{max}-\alpha_{min}$   & 1.437 & 1.677 & 1.865 & 1.461 & 1.335 \\
   $f(\alpha_{min})$   & 0.352 & 0.652 & 0.752 & 0.676 & 0.347 \\
   $f(\alpha_{max})$   & 0.438 & 0.326 & 0.180 & 0.635 & 0.470 \\
   $\Delta f=f(\alpha_{min})-f(\alpha_{max})$   & -0.086 & 0.326 & 0.572 & 0.041 & -0.123 \\ \hline\hline
 \end{tabular}
\end{center}

\section{CONCLUSION}

In summarize, we have investigated the DLA process on a nonuniform
medium which is characterized by a directed 2D SW lattice with two
introduced parameters $(\theta, m)$ which govern the style of the
pattern. It is found that as $m$ increases, the aggregation pattern
become thicker and denser, which indicates the fractal dimension
increases. We also figure out that there exists a maximum value of
$D_0$ in the case of $\theta=0$ for any value of $m$, which implies
the densest aggregation pattern corresponds to the cases of entire
randomly length of long-range connections, neither too long nor too
local. Interestingly, we find that the symmetry of the aggregation
pattern is broken when rewired connections are long enough. The
directed long-range links contribute to the formation of
asymmetrical patterns. The random walk of the particles along the
directed links is irreversible so that the principle of detailed
balance is broken. Hence, the asymmetry of the dynamics finally
results in the asymmetry patterns. To give detailed description of
the asymmetrical pattern, we have performed multifractal analysis on
the patterns. The subtle geometrical difference among these patterns
for different parameter value can be provided by the multifractal
parameters. Although the model we have proposed is very simple, the
simulation results demonstrate that it can capture most of the
general features of asymmetrical growth processes. Other than the
traditional asymmetrical factor such as gravity, magnetic field,
electric field, etc, the asymmetrical factor of our model is the
directed link, which causes the break of dynamics symmetry inherent.
It may be an new explanation for the formation of asymmetrical
fractal behavior in nature.

\end{document}